\documentstyle[11pt]{article}
\newtheorem{theorem}{Theorem}

\newcommand {\ebd} {\stackrel{\Delta} {=}}

\newcommand {\bx} {\mbox{\boldmath $x$}}

\newcommand{\calP}{{\cal P}}

\newcommand{\calS}{{\cal S}}
\newcommand{\calT}{{\cal T}}

\begin{document}
\thispagestyle{empty}
\setcounter{page}{1}
\setlength{\baselineskip}{1.5\baselineskip}
\title{On Context--Tree Prediction of Individual Sequences}
\author{Jacob Ziv and Neri Merhav
\\ \\
Department of Electrical Engineering\\
Technion -- Israel Institute of Technology \\
Haifa 32000, ISRAEL\\
{\tt [jz,merhav]@ee.technion.ac.il}}
\maketitle

\begin{abstract}
Motivated by the evident success of context--tree based methods in lossless data compression,
we explore, in this paper, methods of the same spirit 
in universal prediction of individual sequences.
By context--tree prediction, we refer to a family of prediction schemes, where at each time
instant $t$, after having observed all outcomes of the data sequence $x_1,\ldots,x_{t-1}$, but not
yet $x_t$, the prediction is based on a ``context'' (or a state) that consists of the $k$ most recent
past outcomes $x_{t-k},\ldots,x_{t-1}$, where the choice of $k$ may depend on 
the contents of a possibly longer, though limited, portion of the observed past, 
$x_{t-k_{\max}},\ldots,x_{t-1}$. This is different from the study
reported in \cite{FMG92}, where general finite--state predictors 
as well as ``Markov'' (finite--memory) predictors of fixed order, where studied in
the regime of individual sequences. 

Another important difference between this study and \cite{FMG92} is the asymptotic regime.
While in \cite{FMG92}, the resources of the predictor (i.e., the number of states or the
memory size) were kept fixed regardless of the length $N$ of the data sequence, here we investigate
situations where the number of contexts, or states, is allowed to grow concurrently with $N$.
We are primarily interested in the following fundamental question: What is the critical
growth rate of the number of contexts, below which the performance of the best context--tree
predictor is still universally achievable, but above which it is not? We show that this critical
growth rate is linear in $N$. In particular, we propose a universal context--tree algorithm
that essentially achieves optimum performance 
as long as the growth rate is sublinear, and show that, on the other hand,
this is impossible in the linear case.

\vspace{0.5cm}

\end{abstract}

\section{Introduction}

The problem of universal prediction of 
stochastic processes as well as individual sequences
has received considerable attention throughout the years,
in the literature pertaining to 
a large variety of disciplines, such as information theory, statistics, control
theory, finance, and others (see 
\cite{MF98} for a survey of some
of the results on the theoretical aspects).

In \cite{FMG92}, the problem of universal prediction of individual sequences relative
to the class of finite--state predictors was investigated. 
Given an infinitely long binary sequence
$\bx=(x_1,x_2,\ldots)$, the finite--state predictability, $\pi(\bx)$, was defined as 
\begin{equation}
\pi(\bx)=\lim_{S\to\infty}\limsup_{N\to\infty} \pi_S(x_1,\ldots,x_N),
\end{equation}
where $\pi_S(x_1,\ldots,x_N)$ is the minimum relative frequency of prediction errors achieved among all
finite--state (FS) predictors with no more than $S$ states, when operating on
the first $N$ bits, $x_1,\ldots,x_N$, of the infinite sequence $\bx$. An FS predictor with
$S$ states, or, an {\it $S$--state predictor} for short, is in turn defined by a 
{\it next--state function} $s_{t+1}=g(x_t,s_t)\in\calS$, $|\calS|\le S$,
which recursively updates the state upon receiving a new input, $x_t$, and by an 
{\it output function} $\hat{x}_{t+1}=f(s_t)$,
which provides the prediction of $x_{t+1}$. The main contribution in \cite{FMG92} was in proposing a
universal (randomized) prediction scheme that achieves $\pi(\bx)$ for every $\bx$. This scheme was
based on the incremental parsing procedure of the Lempel--Ziv algorithm \cite{ZL78}. 
Note that since $\pi(\bx)$ is defined by taking the limit of $S\to\infty$ after the limit
supremum over of $N\to\infty$, the regime of the asymptotics dictates that $N$ is very large
compared to $S$.

The present study differs from \cite{FMG92} in two main aspects. The first is that we confine
attention to {\it context--tree} prediction, which means that the current state, $s_t$, does not
necessarily evolve recursively according to a particular next--state function $g$, but may rather
correspond to a certain context, that is, a certain portion of the most recent past
$(x_{t-k},x_{t-k+1},\ldots,x_{t-1})$, where $k$ may vary dynamically 
according to a certain suffix tree, which is subjected to design.
The motivation for exploring context--tree strategies 
stems from their relative simplicity and their success
in lossless data compression applications (see, e.g., 
\cite{MSW04},\cite{STW97},\cite{WS94},\cite{WSS96},\cite{Willems98},\cite{WST95} and references therein).
Quite recently, a context--tree approach was analyzed also in universal prediction
of stochastic processes under certain regularity conditions \cite{JSA02},\cite{Ziv02},\cite{Ziv04}.
Also, as was shown in
\cite{FMG92}, the FS predictability is attainable by finite--memory predictors (also referred to
as ``Markov predictors'' therein), where $k$ is fixed, a--fortiori, it is attainable by the more
general class of context--tree predictors, where $k$ is allowed to vary. 

The second aspect of the
difference between this work and \cite{FMG92} is that here we no longer confine ourselves to the
regime where $N >> S$. By allowing $S$ to grow with $N$ 
at a certain rate, the performance analysis pertaining to
the relative effectiveness of context--tree 
predictors may become more refined and informative in
the sense that it has the potential to 
reveal their advantage over ordinary finite--memory predictors, which under the
regime of \cite{FMG92}, are asymptotically 
as good as general FS predictors anyway, as mentioned above.
Context--tree predictors are intuitively 
superior to finite--memory predictors of fixed order because,
as in data compression,
they allow the flexiblility to allocate more memory resources
(longer contexts) to the ``typical'' patterns, 
that occur more often than others, and less
resources (shorter contexts) to the non--typical ones.

The question that we pose then is the following: 
What is the critical growth rate of
$S=S_N$ as function of $N$, such that below this rate, the asymptotic optimum 
context--tree prediction performance of every 
sequence is still universally achievable, but 
above this rate, it is not? The answer turns out 
to be that this critical rate is {\it linear} in $N$.
More precisely, if $S_N=aN$, ($a$ -- positive constant), 
then no universal predictor (deterministic
or randomized) can attain the optimum 
context--tree prediction performance 
corresponding to $aN$ contexts, simulatenously for all sequences.
Furthermore, for $a=1$, it is easy to show that
the value of this optimum prediction performance (in terms of the
relative error rate)
is zero for any sequence. For a sublinear growth rate
of $S_N$, on the other hand, we propose 
a universal context--based prediction algorithm, whose number of
contexts grows slightly faster than $S_N$, 
and which asymptotically attains the context--tree
predictability pertaining to $S_N$ states, for every $(x_1,\ldots,x_N)$.

The outline of the paper is as follows. 
In Section 2, we give a formal definition of the
problem and state the main result. Sections 3 and 4 are devoted
to proofs.

\section{Problem Formulation and Main Result}

Let $x^N=(x_1,x_2,\ldots,x_N)$, 
$x_t\in\{0,1\}$, $t=1,\dots,N$, designate a binary data sequence
to be sequentially predicted. 
A {\it context--tree predictor} with $S$ contexts (or, with $S$
leaves) is defined as follows. 
The output function, $f(\cdot)$, of the predictor is given by
\begin{equation}
\hat{x}_{t+1}=f(s_t),
\end{equation}
where $\hat{x}_{t+1}\in\{0,1\}$ is the predicted value for $x_{t+1}$ and $s_t$
is the current {\it context} (or, state), 
which takes on values in a finite set $\calS$,
$|\calS|\le S$, $S$ being a positive integer. We allow also randomized 
output functions, namely, random selection of
$\hat{x}_{t+1}\in\{0,1\}$ with respect to (w.r.t.) a conditional probability
distribution given $s_t$. The context 
$s_t$ is determined from the past, $(\ldots,x_{t-1},x_t)$,
by the choice of a {\it context tree},
which is a complete\footnote{By complete binary tree, 
we refer to a binary tree where
every node that is not a leaf has two 
children.} binary tree with $S$ leaves. At time $t$, after having
observed $x_t$, the context $s_t$ is 
determined by reading off the most recent data symbols
in reversed order (first $x_t$, then 
$x_{t-1}$, etc.) and traversing along the
tree according to these symbols, 
starting at the root and ending at a leaf, unless the depth of this leaf is larger than $t$
(which may happen at the beginning of the sequence),
in which case we stop at $x_1$. 
Denoting the resulting depth by
$k=k(\ldots,x_{t-1},x_t)$,
the context will then be given by $s_t=(x_{t-k+1},\ldots,x_t)$.\footnote{
Note that $k$ cannot exceed $S-1$, and so, the context is actually
determined by no more than the $S-1$ most recent symbols.}
Thus, the context--tree is used as a {\it suffix} tree. A context--tree
predictor with $S$ contexts is then defined by a combination of a context--tree 
with a context set $\calS$
and an output function $f:\calS\to\{0,1\}$ (or a set of
conditional distributions $\{P(\cdot|s),~s\in\calS\}$ in the randomized case).
We denote by $\calP_S$ the class of 
all context--tree predictors with $S$ contexts.

Let us now expand the class of predictors $\calP_S$ according to the following model: Given a total
budget of $S$ states, we have the freedom to split it into two subsets of states. One
subset of states, of size $S^C\in\{1,2,\ldots,S\}$, is dedicated to a context--tree of $S^C$
leaves, as before (with $S$ being replaced by $S^C$). The states in this subset will be referred to as
{\it context--tree states}.
The other subset of states,
of size $S^T\le S-S^C$, is dedicated to a finite--state
machine induced by a {\it prefix} tree, 
which is a complete binary tree with a total of $S^T$
nodes (including the root and the internal nodes, 
but not the leaves). The states in this subset will be
referred to as {\it transient states}, and 
each one of the $S^T$ transient 
states corresponds to the root or to an internal node in the prefix tree. The system then works
as follows: It begins at the subset of transient states, and
the initial state, $s_1$, is always the root of the prefix tree. 
As long as $s_t$ is an internal node (or the root) of this tree, the
next state $s_{t+1}=g(x_t,s_t)$ is the 
child of $s_t$ corresponding to the binary value of $x_t$, provided
that this child is an internal node as well, otherwise (i.e., if this
child is a leaf), then
the system passes to the subset of context--tree states,
and then $s_{t+1}$ will be the context 
pertaining to time $t+1$. From this point onward, the system
remains in the subset of context states, 
and operates as described in the previous paragraph.
Thus, the transient states are used only at the beginning of the
sequence, but at certain time $t$ (that may depend on 
the contents of $(x_1,\ldots,x_t)$), there is a transition into the
context state set. We refer to these two modes 
of operation of the system as the {\it transient mode} and
the {\it context--tree mode}, respectively. Let us define $\calP_S^*$ as 
the union, over all pairs of positive integers $\{(S^T,S^C):~ S^T+S^C\le S\}$, of all sets
of combinations of a prefix tree with $S^T$ states and a suffix (context) tree with $S^C$ leaves.
The {\it $S$--th order
context predictability} of $x^N$,
denoted $\kappa(x^N,S)$,
is defined as the minimum
fraction of errors\footnote{When randomized output functions are allowed,
this should be redefined as the
minimum {\it expected} fraction of errors, where
the expectation is w.r.t.\ the randomization. However, it is easy to see
that the best output function is always deterministic.}
achieved over $x^N$ among all
predictors in $\calP_S^*$.

This structure, of a transient mode followed by the context--tree mode, can
be motivated by the following consideration:
Note that in the transient mode, which is active
at the beginning of the sequence, the predictor is
actually using the entire past, $(x_1,\ldots,x_t)$, as its context. This usage of the
entire past can be attributed, in a real-life situation,
to ``training,'' or ``learning.'' During this training time,
in addition to providing predictions, the system ``learns,'' from the whole data available thus far,
what are the ``typical'' patterns
and then, on the basis of this study,
it designs the context--tree predictor to be used in
the context--tree mode, which will remain fixed thereafter.
Since the total memory resources (given by $S$) are limited, they have to be divided
between the training and the size of the context dictionary to be used in the context--tree mode.
Thus, there is a tradeoff, but the definition of the class $\calP_S^*$ allows the full freedom
with regard to the partition between $S^T$ transient states and $S^C$ context--tree states.
On the one extreme, we can take $S^T=0$ and $S^C=S$, which is 
a pure context--tree predictor in $\calP_S$, with no transient mode at all. On
the other extreme, we have $S^T=S-1$ and $S^C=1$, 
where resources are all devoted to the
transient mode, and the context--tree has a root only, which means that
the prediction $\hat{x}_{t+1}$ is constant, independently of past data.

Having defined $\calP_S^*$, let us now allow $S$ grow with $N$, 
and accordingly, redefine the notation
of the total number of states by $S_N$. For a
monotonically non--decreasing sequence 
$\{S_N\}_{N\ge 1}$ of positive integers, we say that
the {\it context predictability is universally
achievable w.r.t.\ $\{S_N\}_{N\ge 1}$} if there exists a randomized
predictor (not necessarily a context predictor), $\hat{x}_t=f_t(x_1,\ldots,x_{t-1})$, $t=1,2,\ldots$,
such that for every infinite sequence $\bx=(x_1,x_2,\ldots)$
\begin{equation}
\label{ua}
\limsup_{N\to\infty}\left[\frac{1}{N}\sum_{t=1}^N \mbox{Pr}\{\hat{x}_t\ne
x_t\}-\kappa(x^N,S_N)\right]\le 0,
\end{equation}
where the probabilities, $\mbox{Pr}\{\hat{x}_t\ne x_t\}$, are w.r.t.\ 
the randomization. We say that a predictor achieves the context
predictability w.r.t.\ $\{S_N\}_{N\ge 1}$ {\it uniformly rapidly} if
the convergence in eq.\ (\ref{ua}) is uniform, i.e.,
\begin{equation}
\limsup_{N\to\infty}\max_{x^N\in\{0,1\}^N}
\left[\frac{1}{N}\sum_{t=1}^N \mbox{Pr}\{\hat{x}_t\ne
x_t\}-\kappa(x^N,S_N)\right]\le 0.
\end{equation}

The questions we address are the following:
\begin{itemize}
\item[1.] What is the fastest growth rate of $\{S_N\}$ such that
the context predictability is still universally
achievable w.r.t.\ $\{S_N\}_{N\ge 1}$ uniformly rapidly?
\item[2.] Whenever the
context predictability is universally
achievable, can we propose a (simple) universal predictor?
\end{itemize}
Theorem 1 answers both questions and tells 
us that this critical growth rate is linear. 

\begin{theorem}
The context predictability w.r.t.\ $\{S_N\}_{N\ge 1}$ is universally
achievable uniformly rapidly if and only if $\lim_{N\to\infty} S_N/N=0$. 
\end{theorem}

\vspace{0.25cm}

\noindent
{\bf Discussion:}
The proof of Theorem 1 consists of the sufficieny part,
where a particular universal (horizon--dependent) predictor is proposed (Section 3)
and the necessity part (Section 4).
As we shall see, the universal predictor proposed in Section 3, bases its
predictions on no more than $2N/M_N$ contexts, where $\{M_N\}_{N\ge 1}$
is a sequence of positive integers tending to infinity such that 
$\lim_{N\to\infty}S_NM_N/N=0$, and so, the number of contexts used
by the algorithm must increase slightly faster than $\{S_N\}$.
As will be seen in Section 3, the best choice of $M_N$, in the sense of minimizing (the upper bound on)
$\max_{x^N\in\{0,1\}^N}
[(1/N)\sum_{t=1}^N \mbox{Pr}\{\hat{x}_t\ne
x_t\}-\kappa(x^N,S_N)]$ is of the order of $(N/S_N)^{2/3}$, which yields a redundancy of
the order of $(S_N/N)^{1/3}$. It should be noted that it is also possible to obtain a redundancy
rate of $O((S_N\log S_N)/N)$, which may be better in some cases, by
using the {\it expert--advice} methodology (cf.\ the relevant
references in \cite{MF98}), where the
``experts'' are all the members of $\calP_{S_N}^*$.
However, the implementation of the 
expert--advice algorithm is extremely complex because it
needs to apply all predictors of $\calP_{S_N}^*$ in parallel.
The proposed horizon--dependent 
algorithm is next modified to be horizon--independent.

As for the necessity part of Theorem 1, 
we assume that $S_N=aN+1$ for some positive constant $a\le 1$,
and demonstrate that there is 
a set of sequences $\{x^N\}$ for which, on the one hand,
$\kappa(x^N,aN+1)=0$, but on the other hand, for every 
universal predictor (which may be deterministic or randomized, and with
unlimited resources), at least one of 
these sequences would yield no less than $aN/2$ errors.
Stated in the mathematical language, we have:
\begin{equation}
\max_{x^N\in\{0,1\}^N}
\left[\frac{1}{N}\sum_{t=1}^N \mbox{Pr}\{\hat{x}_t\ne
x_t\}-\kappa(x^N,aN+1)\right]\ge \frac{a}{2}
\end{equation}
for all $N$, and so, when $\lim_{N\to\infty}S_N/N =a > 0$, the context predictability
is not universally achievable uniformly rapidly. The question of universal achievability
which is not uniformly rapid, in the linear case, remains open.

\section{A Universal Prediction Scheme -- Proof of Sufficiency}

For a given $N$, choose a positive integer $M_N$, and
consider the following recursive definition of {\it prediction context},
which also defines the proposed algorithm.

Let $k_0=k_0(x_1,\ldots,x_t)$ denote 
the largest positive integer $k$ such that
the following two conditions hold at the same time:
\begin{itemize}
\item [1.] The string $(x_{t-k+1},\ldots,x_t)$ appears 
(possibly, with overlaps) at least $M_N$ times along
$(x_1,\ldots,x_t)$. 
\item [2.] The string $(x_{t-k+2},\ldots,x_t)$ has already been
used as the {\it prediction context} at least $M_N$ times in the past.
\end{itemize}
If no such $k$ exists, define $k_0=0$.
The string $(x_{t-k_0+1},\ldots,x_t)$
is referred to as the {\it prediction context} used at time $t$, and
in the case $k_0=0$, the context $s_t$ is defined as ``null,'' i.e., ``no context.''

Next, consider the prediction scheme of \cite{FMG92}, defined w.r.t.\
the prediction context $s_t=(x_{t-k_0+1},\ldots,x_t)$.
In particular, at each time instant $t$, determine the context using
the above described rule, and randomly draw the 
prediction $\hat{x}_{t+1}$ according
to the conditional distribution 
$p_t(\hat{x}_{t+1}=1|s_t)=\phi(\hat{p}_t(1|s_t),N(s_t))$, 
where $\phi$ is defined as follows:
\begin{equation}
\label{output}
\phi(\alpha,n)
=\left\{\begin{array}{ll}
0 & \alpha < \frac{1}{2}-\epsilon_n \\
\frac{1}{2\epsilon_n}(\alpha-\frac{1}{2})+\frac{1}{2} &
\frac{1}{2}-\epsilon_n \le \alpha\le
\frac{1}{2}+\epsilon_n\\
1 & \alpha > \frac{1}{2}+\epsilon_n \end{array}\right.
\end{equation}
with $\epsilon_n\ebd 1/(2\sqrt{n+2})$, and
where $\hat{p}_t(1|s)=[N_t(s,1)+1/2]/[N_t(s)+1]$, $N_t(s)$ being the number of
occurrences of the context $s$ (w.r.t.\ the above rule) along
$(x_1,\ldots,x_{t-1})$
and $N_t(s,1)$ is the number of times these
appearances of context $s$ were followed by ``1''.

We next analyze the performance of this prediction scheme in
comparison to the best reference predictor in $\calP_{S_N}^*$, with a set $\calS_N^T$ of
$S_N^T$ transient states, and a set $\calS_N^C$ of
$S_N^C$ context states, $S_N^T+S_N^C\le S_N$. An upper bound on the redundancy,
$[(1/N)\sum_{t=1}^N \mbox{Pr}\{\hat{x}_t\ne
x_t\}-\kappa(x^N,S_N)]$, will be obtained by bounding $(1/N)\sum_{t=1}^N \mbox{Pr}\{\hat{x}_t\ne
x_t\}$ from above, and bounding $\kappa(x^N,S_N)$ from below. We begin with the latter by counting
only errors that occur during the context--tree mode of the reference predictor, 
which lasts at least $N-S_N^T$ time units, as the
transient mode cannot last longer than $S_N^T$ instants.
For the given $x^N$, let $(s_1,\ldots,s_N)$
be the sequence of states that would 
have been obtained had {\it only} the context--tree
machine of the reference predictor been used, from $t=1$ to $t=N$.
As is shown in \cite{FMG92}, the number 
of errors made by such a (pure context--tree) predictor
is given by
$\sum_{s\in\calS_N^C}\min\{N(s,0),N(s,1)\}$,
where $N(s,x)$, $s\in\calS_N^C$, $x\in\{0,1\}$,
is the number of joint occurrences of $s_t=s$ and $x_{t+1}=x$
along the pair of sequences $(s^N,x^N)$. The joint count of $s_t=s$ and $x_{t+1}=x$, during 
the context--tree mode only, cannot then be smaller than $N(s,x)-S_N^T$, and so,
\begin{eqnarray}
\kappa(x^N,S_N)&\ge&\frac{1}{N}\left[\sum_{s\in\calS_N^C}\min\{N(s,0),N(s,1)\}-S_N^T\right]\nonumber\\
&\ge&\frac{1}{N}\left[\sum_{s\in\calS_N^C}\min\{N(s,0),N(s,1)\}-S_N\right].
\end{eqnarray}
As was also shown in \cite{FMG92}, when the predictor (\ref{output})
is applied, the contribtution of each state $s$ to the expected number
of prediction errors, 
$EN_e(s)\ebd\sum_{t:s_t=s}\mbox{Pr}\{\hat{x}_t\ne x_t\}$, is upper bounded by
\begin{equation}
\label{seq}
EN_e(s)\le \min\{N(s,0),N(s,1)\}+\sqrt{N(s)+1}+\frac{1}{2},
\end{equation}
where $N(s)=N(s,0)+N(s,1)$ is the number of occurrences of $s$.

Consider the above described universal prediction scheme applied
to $x^N$, and let us denote now the sequence of contexts, generated by
this algorithm, as $\hat{s}^N=(\hat{s}_1,\ldots,\hat{s}_N)$ (to distinguish
from the contexts of the context--tree component of the reference predictor of $\calP_{S_N}^*$), and
let $\hat{\calS}_N$ denote the set of contexts generated this way.

We first observe that there are at most $2M_NS_N^C$ times instants
where $\hat{s}_t$ is a suffix of $s_t\in\calS_N^C$. This follows from the following
consideration. In a full binary tree with $S_N^C$ leaves, like the
tree corresponding to the reference predictor,
there are always $S_N^C-1$ internal nodes (including the root), pertaining
to all possible states which are suffixes of some state in $\calS_N^C$.
Now, by construction of the algorithm, every such internal node $s'$ is
used as a prediction context no more than $2M_N$ times. This is because upon
the $(2M_N+1)$--st time, either the pattern $(0,s')$ or $(1,s')$ has appeared
at least $M_N$ times, and thus both conditions for extending the prediction context
by one bit are satisfied. Thus, the total number of times that suffixes of
contexts in $\calS_N^C$ are used as prediction contexts cannot exceed
$2M_N(S_N^C-1)$. We will further upper 
bound this number by $2M_NS_N$, for simplicity.

In the remaining time instants, of course, 
either $\hat{s}_t=s_t$ or $s_t$ becomes a suffix of $\hat{s}_t$.
Correspondingly, for a given $s\in\calS_N^C$,
let $\calT_s$ denote the sub--tree of prediction contexts,
rooted at $s$, that are generated by the algorithm, i.e., all generated
contexts $\{\hat{s}\}$ suffixed by $s$ (including 
$s$ itself as the root). Following eq.\ (\ref{seq}), the expected
number of errors is bounded by
\begin{equation}
\label{1st}
\frac{1}{N}\sum_{t=1}^N\mbox{Pr}\{\hat{x}_t\ne x_t\}
\le 2M_NS_N+\sum_{s\in\calS_N^C}
\sum_{\hat{s}\in \calT_s} \left[\min\{N(\hat{s},0),N(\hat{s},1)\}
+\sqrt{N(\hat{s})+1}+\frac{1}{2}\right],
\end{equation}
where the first term, $2M_NS_N$, accounts for worst case of totally erroneous prediction
at all $2M_NS_N$ visits at states $\{\hat{s}\}$ that are 
suffixes of some states in $\calS_N^C$, 
and the second term is an upper bound 
on the expected number of errors at all other times.
Now, let us decompose the second term into
\begin{equation}
A\ebd\sum_{s\in\calS_N^C}
\sum_{\hat{s}\in \calT_s} \min\{N(\hat{s},0),N(\hat{s},1)\}
\end{equation}
and
\begin{equation}
B\ebd\sum_{s\in\calS_N^C}
\sum_{\hat{s}\in \calT_s}
\left[\sqrt{N(\hat{s})+1}+\frac{1}{2}\right].
\end{equation}
We shall now bound each one of them separately. As for $A$, we have
\begin{eqnarray}
\label{Aterm}
A&\le&\sum_{s\in \calS_N^C}
\min\left\{\sum_{\hat{s}\in \calT_s} N(\hat{s},0),
\sum_{\hat{s}\in \calT_s} N(\hat{s},1)\right\}\nonumber\\
&\le&
\sum_{s\in\calS_N^C}\min\{N(s,0),N(s,1)\}\nonumber\\
&\le&
N\cdot\kappa(x^N,S_N)+S_N.
\end{eqnarray}
Regarding $B$, we have the following consideration: As mentioned earlier,
for internal nodes in $\calT_s$ (and a--fortiori for the leaves), we know that
$N(\hat{s})$ cannot exceed $2M_N$, and so,
\begin{equation}
B\le\sum_{s\in\calS_N^C}\sum_{\hat{s}\in \calT_s}
\left(\sqrt{2M_N+1}+\frac{1}{2}\right)
=\left(\sqrt{2M_N+1}+\frac{1}{2}\right)\cdot\sum_{s\in\calS_N^C}|\calT_s|.
\end{equation}
Now, $\sum_{s\in\calS_N^C}|\calT_s|$ 
is of course, upper bounded by
the total number of contexts generated by the proposed universal predictor. As
every internal node of the context--tree generated appears at least $M_N$
times (by the second condition 
that defines the algorithm), the total number of internal nodes 
of $\hat{\calS}_N$ cannot exceed $N/M_N$, and so,
the total number of nodes (including the leaves) cannot exceed $2N/M_N+1$.
Thus, $\sum_{s\in\calS_N^C}|\calT_s|\le 2N/M_N+1$, 
and we can further upper bound $B$ by
\begin{equation}
B\le\left(\sqrt{2M_N+1}+\frac{1}{2}\right)\cdot\left(\frac{2N}{M_N}+1\right),
\end{equation}
which upon normalizing by $N$ becomes
\begin{equation}
\frac{B}{N}\le \left(2\sqrt{\frac{2}{M_N}+
\frac{1}{M_N^2}}+\frac{1}{M_N}\right)\cdot\left(1+\frac{M_N}{2N}\right).
\end{equation}
The total expected excess frequency of errors (redundancy) is thus
\begin{eqnarray}
\frac{1}{N}\sum_{t=1}^N\mbox{Pr}\{\hat{x}_t\ne x_t\}
-\kappa(x^N,S_N)&\le&
\left(2\sqrt{\frac{2}{M_N}+
\frac{1}{M_N^2}}+\frac{1}{M_N}\right)
\cdot\left(1+\frac{M_N}{2N}\right)+\nonumber\\ 
& &\frac{(2M_N+1)S_N}{N},
\end{eqnarray}
where the additional term comes from the first term of the r.h.s.\
of eq.\ (\ref{1st}) and the right--most side of eq.\ (\ref{Aterm}). The conditions for vanishing
redundancy are then $M_N\to\infty$ 
and $M_NS_N/N\to 0$. Both conditions can be satisfied at
the same time as long as $S_N$ is sublinear in $N$. As 
the r.h.s.\ is independent of $x^N$, the convergence to zero is uniformly
fast. This completes the proof of the sufficiency part. $\Box$

Two comments are in order at this point:
\begin{itemize}
\item [1.] Note that the asymptotically optimum growth rate of $M_N$
(in the sense of minimizing the r.h.s.) is $M_N=O((N/S_N)^{2/3})$,
which yields $B/N\le O((S_N/N)^{1/3})$. 
\item [2.] The above algorithm is horizon--dependent, i.e., the length of the sequence, $N$, has to
be known ahead of time in order to determine the value of $M_N$. It is not difficult, however, to
modify this algorithm so as to be horizon--independent. One way to do that is the following: Instead
of defining the required number of context repetitions, in conditions 1 and 2 of the algorithm, to depend directly
on $N$, let us define it as depending on $k$, the length of the examined context. 
More specifically, let us replace
$M_N$ by $M(k)$ and by $M(k-1)$ in conditions 1 and 2, respectively, 
where $\{M(k)\}_{k\ge 1}$ is a certain monotonic sequence 
of positive integers that tends to infinity. The reader is referred to the appendix for more details
on the redundancy analysis and the considerations regarding the choice of the sequence $\{M(k)\}$.
It is also demonstrated, in the appendix, that the (upper bound on the) redundancy term of this algorithm
decays faster than that of the LZ--based algorithm proposed in \cite{FMG92}.
\end{itemize}

\section{Proof of Necessity} 

Let $a\in(0,1]$ be given, and let $S_N=aN+1$, 
assuming without essential loss of generality
that $aN$ is integer.
Consider the recursive generation of a sequence $x^N$ by
$x_t=f(s_t)$, $t=1,2,...,N$, where $s_t$ is the state
associated with previously generated symbols, and $f$ is the output function, corresponding
to a certain member in $\calP_{aN+1}^*$. Clearly,
when this predictor is applied to the very same sequence
that it has generated, then there are no prediction errors, and so,
$\kappa(x^N,aN+1)=0$ for every such sequence.

Next, consider a subset of $2^{aN}$ pure 
transient--state predictors from $\calP_{aN+1}^*$, i,e., 
predictors with $S_N^T=aN$ and $S_N^C=1$,
whose associated $x$-sequences (generated
as above) start with {\it all} $2^{aN}$ possible binary strings of length
$aN$ correspondingly. That is, the first predictor generates a sequence
that begins with $aN$ zeroes, the second predictor generates a 
sequence whose first $aN$ bits are $(0,0,...,0,1)$,
and so on. Clearly, there are enough degrees
of freedom to do that: Given any desired binary 
string $(x_1,\ldots,x_{aN})$ of the first $aN$ bits of $x^N$,
consider the finite--state (transient) machine corresponding
to a prefix tree whose
internal nodes are $\emptyset$ (the null string),
$\{x_1\}, \{x_1,x_2\},\ldots,\{x_1,...,x_{aN}\}$, and 
whose leaves are
$\{\bar{x}_1\}, \{x_1,\bar{x}_2\},\{x_1,x_2,\bar{x}_3\},\ldots,
\{x_1,x_2,\ldots,x_{aN-1},\bar{x}_{aN}\}$, 
$\bar{x}_i$ being the complement of $x_i$, $i=1,\ldots,aN$.
Now, apply to each of the internal nodes an output function that will give
the next desired outcome, i.e., 
$f(\emptyset)=x_1$, $f(\{x_1\})=x_2$, 
$f(\{x_1,x_2\})=x_3,\ldots, f(\{x_1,x_2,\ldots,x_{aN-1}\})=x_{aN}$.
This construction guarantees that each one of the 
$2^{aN}$ context--tree predictors will generate a different sequence
because all these sequences differ from each other even in their
first $aN$ bits. 

Finally, define a random vector $X^N$, which is distributed
uniformly across all these $2^{aN}$ $N$--vectors.
Now, for any randomized
predictor, with no matter how many states, the expected
fraction of errors (where the expectation is
both w.r.t.\ the ensemble of $X^N$ and w.r.t.\ possible randomization)
is lower bounded as follows:
\begin{equation}
\frac{1}{N}\sum_{t=1}^N\mbox{Pr}\{\hat{x}_t\ne X_t\}
\ge \frac{1}{N}\sum_{t=1}^{aN}\mbox{Pr}\{\hat{x}_t\ne X_t\}=\frac{a}{2},
\end{equation}
where the last equality is due to the fact that $(X_1,\ldots,X_{aN})$
is, in fact, governed by the memoryless
binary symmetric source (independent, fair
coin tosses) since the distribution is uniform over all $2^{aN}$ strings
on length $aN$. Clearly, every predictor makes exactly 50\% errors on
the binary symmetric source. 
It therefore follows that for any randomized predictor, there exists at least
one vector $x^N$, out of the above defined ensemble of $2^{aN}$ vectors,
for which the expected fraction of errors is not below $a/2$.
This completes the proof of the necessity part. 

Note that for the case $a=1$,
we have $\kappa(x^N,N+1)=0$ for {\it every} sequence, but any predictor would
perform at least as bad as random guessing (50\% errors) on some sequence.

\section*{Appendix}
\renewcommand{\theequation}{A.\arabic{equation}}
   \setcounter{equation}{0}

In this appendix, we show how the performance analysis of Section 3 should be
modified if the horizon--dependent algorithm is replaced by the
the horizon--independent algorithm described in the second comment at the end
of Section 3.

In analogy to eq.\ (\ref{1st}), we have two main redundancy terms:  The first
term is  the summation of $2M(d_s)$ 
over all internal nodes $\{s\}$ of the context--tree $\calS_N^C$ (replacing the term $2M_NS_N$), where $d_s$
stands for the depth of state $s$ in the context--tree, i.e., the distance from
of $s$ from the root. This term is further bounded by $2S_N\max_{s\in\calS_N^C}M(d_s)
=2S_NM(\max_{s\in\calS_N^C}d_s)\le 2S_NM(S_N)$, where we have used the fact that
the deepest leaf in a compete tree with $S_N$ leaves cannot be more than $S_N$ branches away
from the root. The second term is $B$, which is now upper bounded as follows:
\begin{eqnarray}
B&=&\sum_{s\in\calS_N^C}\sum_{\hat{s}\in\calT_s}\left(\sqrt{N(\hat{s})+1} +\frac{1}{2}\right)\nonumber\\
&=&\sum_{s\in\calS_N^C}|\calT_s|\sum_{\hat{s}\in\calT_s}
\frac{1}{|\calT_s|}\cdot\left(\sqrt{N(\hat{s})+1} +\frac{1}{2}\right)\nonumber\\
&\le&\sum_{s\in\calS_N^C}|\calT_s|\left(\sqrt{\frac{1}{|\calT_s|}\sum_{\hat{s}\in\calT_s}
N(\hat{s})+1} +\frac{1}{2}\right)\nonumber\\
&\le&\sum_{s\in\calS_N^C}|\calT_s|\left(\sqrt{\frac{N(s)}{|\calT_s|}
+1} +\frac{1}{2}\right)\nonumber\\
&=&\sum_{s\in\calS_N^C}\sqrt{|\calT_s|}\cdot\sqrt{
N(s)+|\calT_s|} +\frac{1}{2}\sum_{s\in\calS_N^C}|\calT_s|\nonumber\\
&\le&\sqrt{\sum_{s\in\calS_N^C}|\calT_s|\cdot\sum_{s\in\calS_N^C}[N(s)+|\calT_s|]}+ 
\frac{1}{2}\sum_{s\in\calS_N^C}|\calT_s|\nonumber\\
&\le&\sqrt{\sum_{s\in\calS_N^C}|\calT_s|\cdot(N+\sum_{s\in\calS_N^C}|\calT_s|)}+ 
\frac{1}{2}\sum_{s\in\calS_N^C}|\calT_s|,
\end{eqnarray}
where the first inequality follows from the concavity of the square--root function, and
the second to the last inequality follows from the Schwartz--Cauchy inequality.
Now, $\sum_{s\in\calS_N^C}|\calT_s|$, which is upper bounded by the total number of
contexts generated by the algorithm, $|\hat{\calS}_N|$, is in turn, upper bounded by the following consideration:
Denoting by $\tilde{\calS}_N$,
the set of internal nodes of $\hat{\calS}_N$, we have for every positive integer $j$:
\begin{eqnarray}
N&\ge&\sum_{\hat{s}\in\tilde{\calS}_N}M(d_{\hat{s}})\nonumber\\
&\ge&\sum_{\hat{s}\in\tilde{\calS}_N:~d_{\hat{s}}\ge j}M(d_{\hat{s}})\nonumber\\
&\ge&\sum_{\hat{s}\in\tilde{\calS}_N:~d_{\hat{s}}\ge j}M(j)\nonumber\\
&\ge&(|\tilde{\calS}_N|-2^j+1)\cdot M(j),
\end{eqnarray}
where we have used the fact that the number of nodes with depth less than $j$ cannot exceed
$\sum_{i=0}^{j-1}2^i=2^j-1$. We therefore have
\begin{equation}
|\tilde{\calS}_N|\le 2^j-1+\frac{N}{M(j)},
\end{equation}
and so,
\begin{equation}
\sum_{s\in\calS_N^C}|\calT_s|\le |\hat{\calS}_N|< 2^{j+1}+\frac{2N}{M(j)},
\end{equation}
which follows from the fact that in a complete binary tree with $m$ internal nodes,
the total number of nodes is $2m+1$.
Since this is true for every $j$, we can take the minimum over $j$.
Let us then denote
\begin{equation}
\psi(N)=\frac{1}{N}\min_j\left(2^{j+1}+\frac{2N}{M(j)}\right).
=2\min_j\left(\frac{2^j}{N}+\frac{1}{M(j)}\right).
\end{equation}
We therefore obtain the following upper bound to the redundancy:
\begin{equation}
\frac{1}{N}\sum_{t=1}^N\mbox{Pr}\{\hat{x}_t\ne x_t\}-\kappa(x^N,S_N)\le
\frac{2S_N(M(S_N)+1)}{N}+\sqrt{\psi(N)[1+\psi(N)]}+\frac{\psi(N)}{2}.
\end{equation}
The guidelines regarding the choice of the sequence $\{M(k)\}$ are, in principle, aimed at minimizing
the r.h.s.\ of the last inequality. Obvioulsy, the faster is the growth rate of $\{M(k)\}$,
the faster $\psi(N)$ decays, but on the other hand, the first term 
above is enlarged. Moreover, this dictates
an interesting tradeoff with regard to universal achievability. If one wishes to compete with
the context predictability for every sublinear growth rate of $\{S_N\}$, then $M(k)$ should be a constant $M_0$
(otherwise $S_NM(S_N)/N$ may not tend to zero), but then $\psi(N)$ tends to a constant, which
can be made arbitrarily small for large enough $M_0$. Thus, the context predictability is achieved
within an arbitrarily small $\epsilon > 0$,
but not strictly achieved. If, on the other hand, one is somewhat less ambitious, and is only interested
in achieving the context predictability for slower sequences $\{S_N\}$, i.e., those for which
$\{S_NM(S_N)/N\}$ still vanishes for a certain choice of the sequence $\{M(k)\}$, then this
is accomplished by the algorithm. 
For example, if $M(k)=2^k$, then $\psi(N)=O(1/\sqrt{N})$, 
but then $\{S_N\}$ of the reference class is only allowed to grow
slower than logarithmically in $N$, for the purpose of comparison.

Finally, it is interesting to compare the performance of the proposed
horizon--independent algorithm to that of the LZ--based algorithm of \cite{FMG92}.
To this end, let us even assume that $S_N=S=2^k$ is fixed (not growing with $N$),
and that our reference predictor is a pure context--tree algorithm (with no transient states),
where the context--tree is the full binary tree whose leaves are all the $2^k$ binary $k$--tuples,
in other words, a finite--memory (``Markov'') predictor of order $k$. In \cite[Theorem 4]{FMG92},
it is asserted that the (upper bound on the) redundancy 
of the LZ--based predictor w.r.t.\ this finite--memory predictor decays
at the rate of $1/\sqrt{\log N}$. Here, on the other hand, if we choose, for example, $M(k)=2^k$, as
suggested above, then the redundancy would decay at the rate of $N^{-1/4}$, which is better.
Moroever, the choice $M(k)=2^k$ may not even be the best possible choice. One can come close
to the rate of $N^{-1/2}$ by letting $\{M(k)\}$ grow sufficiently rapidly.

\end{document}